\documentclass[%
reprint,
 amsmath,amssymb,
 aps,
 prb,
 unsortedaddress,
 superscriptaddress
]{revtex4-2}

\usepackage{dcolumn}
\usepackage{bm}
\usepackage[usenames,dvipsnames]{xcolor}
\usepackage[caption=false]{subfig}
\usepackage{natbib}
\usepackage{caption}
\usepackage{baskervald}
\usepackage{graphicx}
\usepackage{floatrow}
\usepackage[hidelinks]{hyperref}
\usepackage{amsmath, mathtools}

\newcommand{\eqsplit}[1]{\begin{equation}\begin{split}#1\end{split}\end{equation}}

\newcommand{\llangle}{\langle\langle}
\newcommand{\rrangle}{\rangle\rangle}
\newcommand{\occ}{\mathrm{occ}}

\newcommand{\dchemistry}{Department of Chemistry, Rice University, Houston, TX 77005-1892}

\begin{document}
\preprint{APS/123-QED}

\title{Electron localization in disordered quantum systems at finite temperatures}

\author{Chong Sun}
 \affiliation{\dchemistry}
\email{sunchong137@gmail.com}

\date{\today}

\begin{abstract}
We study electron localization in disordered quantum systems, focusing on both individual eigenstates and thermal states. We employ complex polarization as a numerical indicator to characterize the system's localization length. Furthermore, we assess the efficacy of mean-field approximation in providing a quantitative analysis of such systems. Through this study, we seek to provide insight into the following aspects: the behavior of electron localization as a function of interaction, disorder, and temperature; whether thermal states and highly excited states exhibit similar properties in many-body localized systems; and the reliability of the mean-field approximation in weak-interaction scenarios.
\end{abstract}

\maketitle

\section{Introduction\label{sec:intro}}
Many-body localization (MBL) in disordered quantum systems with interactions~\cite{Gornyi2005Interacting, Basko2006Metal, Basko2006Problem, Rahul2015Many, Abanin2017Recent, Vu2022Fermionic} has drawn much attention due to the potential of preserving symmetry-broken orders and deferring quantum decoherence~\cite{Huse2013Localization, Parameswaran2018Many}. In such systems, disorder reduces crosstalking among subsystems, and thus hinders quantum state thermalization~\cite{Rigol2008Thermalization, Kaufman2016Quantum, Deutsch2018Eigen}. 
Theoretical characterizations of MBL include a Poissonian distribution of the ratio of adjacent energy gaps~\cite{Ogsanesyan2007Localization}, area-law and logarithm-growth of the entanglement entropy~\cite{Bardarson2012Unbounded, Bauer2013Area, Serbyn2013Universal}, presence of quasi-local integrals of motion (LIOMs)~\cite{Serbyn2013Local}, etc. Notably, MBL emerges not only in the ground state but across all eigenstates. This suggests that MBL can be viewed as a form of localization in the Fock space, where transitions to proximate eigenstates are impeded.

Conflicting conclusions have been made in studies of the dc conductivity of interacting disordered systems at finite temperatures~\cite{Karahalios2009Finite, Berkelbach2010Conductivity}, where a nonzero conductivity persists at high temperatures regardless of the disorder strength. However, these results are not necessarily at odds with established understandings. First, traditional MBL arguments typically consider an isolated system, where thermalization follows microcanonical statistics, in contrast to the canonical or grand canonical frameworks often applied in finite-temperature simulations. Second, because MBL entails localization within the Fock space, it is expected that individual eigenstates exhibit localization characteristics, which may not be reflected in the thermal average across multiple eigenstates.
Nevertheless, exploring disordered quantum systems at finite temperatures remains a valuable pursuit, given the practical challenges of achieving complete isolation in real-world systems.

In this study, we investigate real-space electron localization within disordered quantum systems, employing the {complex polarization operator}, a concept introduced in a series of works by Resta and Sorella in 1990s~\cite{Resta1992Theory, Resta1993Macroscopic, Resta1998Quantum, Resta1999Electron, Resta2002Why}. 
The recently proposed "periodic position operator"~\cite{Valenca2019Simple, Evangelisti2022Unique} and "imaginary vector potential"~\cite{Heussen2021Extracting} adopted similar concepts. 
We provide formulations for evaluating the expectation values of the complex polarization for both individual eigenstates and thermal states.
We demonstrate that full electron localization persists in any given eigenstate of the system under strong disorder. 
At high temperatures, however, electrons exhibit a large spread length even under strong disorder, aligning with previous findings on finite-temperature conductivity~\cite{Karahalios2009Finite, Berkelbach2010Conductivity}. 
We also assess the efficacy of mean-field simulations for such systems under weak interactions. Utilizing the Hartree-Fock (HF) method as our mean-field framework, we
demonstrate that the HF approach provides reliable quantitative descriptions of electron localization in conditions of strong disorder. 

In Section~\ref{sec:compol}, we introduce the complex polarization operator and mathematically connect it to the electron localization length. In Section~\ref{sec:niform} and ~\ref{sec:iform}, we provide the ground-state and finite-temperature formulations to evaluate electron localization length for both Slater determinants and correlated quantum states. In Section~\ref{sec:results}, we conduct HF and full-configuration interaction (FCI) calculations on a one-dimensional disordered model, and characterize the effects of disorder strength, temperature, and system size on electron localization. We conclude in Section~\ref{sec:conclusion} with a discussion of the broader implications of our numerical findings for disordered quantum systems.

\section{Theory\label{sec:theory}}
\subsection{Electron localization\label{sec:compol}}
The electron localization length $\lambda$ in a one-dimensional system can be described by the electron quadratic spread
\eqsplit{
\lambda^2 = \langle x^2\rangle - \langle x\rangle^2,
}
where $x$ is the position operator of an electron. The localization length $\lambda$ has a finite value for a localized state and diverges for a delocalized state.
This divergence presents practical challenges for numerical simulations. In addition, the electron localization length in a system with periodic boundary conditions (PBC) becomes ambiguous. 
To address the above issues, Resta and Sorella introduced the \textit{complex polarization operator} $Z$, which relates to the electron quadratic spread by~\cite{Resta1998Quantum, Resta1999Electron}
\eqsplit{\label{eq:quad_spread}
\lambda^2
 = -\frac{L^2}{2\pi^2} \log |Z|,
}
where the magnitude of $Z$ satisfies $0 \leq |Z| \leq 1$. The electrons are fully diffused when $|Z| = 0$ and fully localized when $|Z| = 1$. This relation is still valid at finite temperatures.

\begin{figure}
    \centering
    \includegraphics[width=0.9\linewidth]{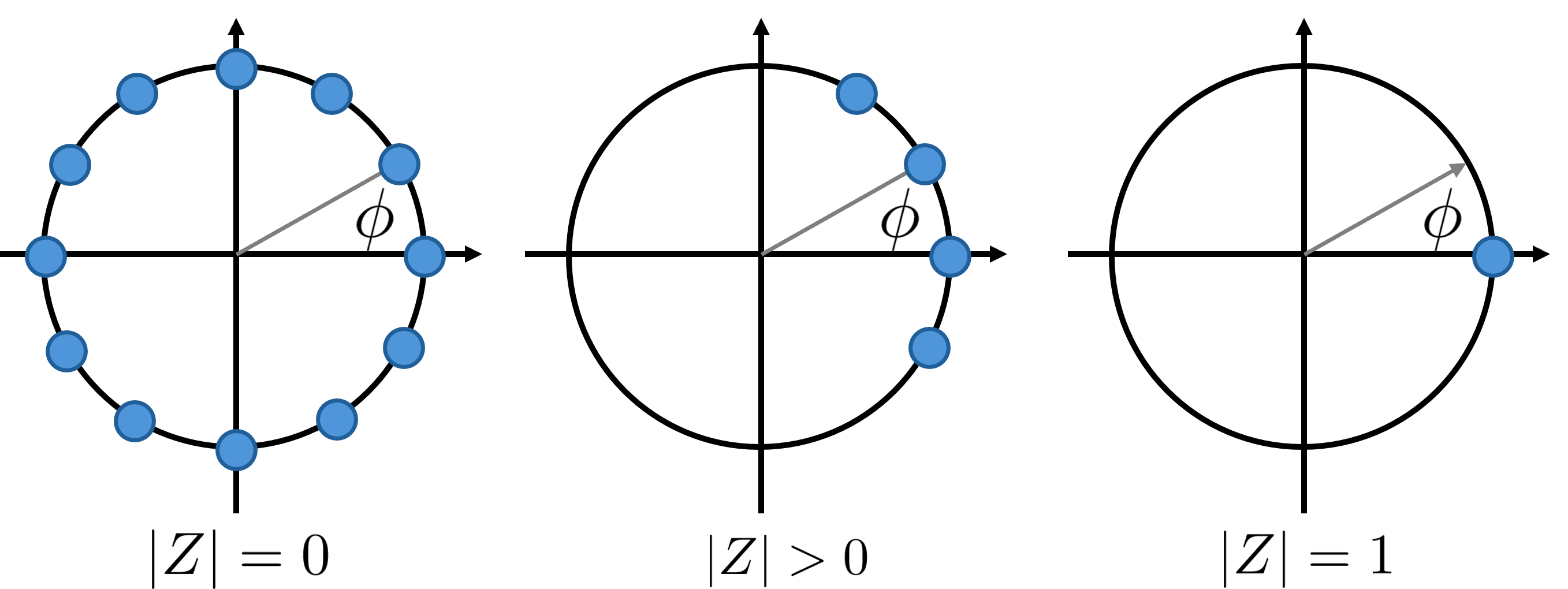}
    \caption{Connection between electron localization length and the magnitude of the complex polarization $|Z|$. The position of an electron on the one-dimensional lattice is mapped to an angle $\phi$ ranging from $0$ to $2\pi$.}
    \label{fig:fig0}
\end{figure}

Herein, we introduce the complex polarization operator for a one-dimensional lattice with PBC as 
\eqsplit{\label{eq:compol_1d}
Z = \exp\left(i\frac{2\pi X}{L}\right), \quad {X} = \sum_{j=1}^{N} {x}_j,
}
where $x_j$ is the position operator of the $j$th electron, $L$ is the lattice length, and $N$ is the total number of electrons. The expectation value of $Z$, denoted as $\langle Z\rangle = |Z|e^{i\gamma}$, is a complex number, where $\gamma$ is related to the single-point Berry phase~\cite{Resta2000Berry}. Note that while $X$ is a one-body operator, $Z$ serves as an $N$-body operator, capturing the collective behavior of all electrons. Fig.~\ref{fig:fig0} qualitatively illustrates the relationship between $|Z|$ and the electron localization length, where the position of an electron on the one-dimensional lattice is mapped to an angle ranging from $0$ to $2\pi$.
When an electron can occupy all possible positions, $\langle Z\rangle = 0$ due to the phase cancellation. Conversely, complete localization of the electron results in $\langle Z\rangle = e^{i\gamma_0}$ and $|Z| = 1$. 
Generalization to the multi-dimensional formulation of Eq.~\eqref{eq:compol_1d} is simply the direct product of $Z$ in each dimension.

\subsection{Non-interacting formulation\label{sec:niform}}
Although $Z$ is a complex $N$-body operator, its application to a Slater determinant produces another Slater determinant, as a result of Thouless theorem~\cite{Thouless1960Stability, Rosensteel1981Nondeterminantal, Ripka1986Quantum}. 
This property facilitates the evaluation of $\langle Z\rangle$ when the system is non-interacting and can be described by a Slater determinant $|\Phi\rangle$, reducing the problem to calculating the overlap between two Slater determinants
\eqsplit{\label{eq:expectation_slater}
\langle \Phi| {Z} |\Phi\rangle = \langle \Phi|\tilde{\Phi}\rangle = \det\left[\mathbf{C}_{\occ}^\dagger \tilde{\mathbf{C}}_{\occ}\right],
}
where $|\tilde{\Phi}\rangle = Z|\Phi\rangle$, and the columns of $\mathbf{C}_{\occ}$ and $\tilde{\mathbf{C}}_{\occ}$ denote the occupied orbital coefficients on which $|\Phi\rangle$ and $|\tilde{\Phi}\rangle$ are constructed, respectively. We have assumed that $|\Phi\rangle$ is normalized. 
If the position/site basis is used, one can derive an $L\times L$ matrix representation of $Z$ as
\eqsplit{\label{eq:matrix_compol}
\langle \mu |Z|\nu\rangle = \exp\left(i\frac{2\pi x_\mu}{L}\right)\delta_{\mu\nu},
}
where $\{|\mu\rangle\}$ is the site basis. Therefore, under the site basis, $\mathbf{Z}$ is a diagonal matrix and $\tilde{\mathbf{C}} = \mathbf{ZC}$.
The derivations of Eq.~\eqref{eq:expectation_slater} and \eqref{eq:matrix_compol} are provided in Appendix~\ref{sec:apdx_slater_thouless}.

We now turn to the finite-temperature formulations for non-interacting systems, derived using the thermofield theory~\cite{Umezawa1982thermo, Sun2021Finite}. The thermal average of the complex polarization $Z$ is given by
\eqsplit{\label{eq:ft_cplx_mf}
\langle Z\rangle (\beta) &= \frac{\text{det}\left[\mathbf{Z}\bm{\rho}(\beta) + \mathbf{I}\right]}
{\text{det}\left[\bm{\rho}(\beta) + \mathbf{I}\right]}
,\\
\bm{\rho}(\beta) &= \exp[-\beta (\mathbf{h} -\mu)]
}
where $\beta = 1/k_B T$ is the inverse temperature,  $\bm{\rho}(\beta)$ is the density matrix, $\mathbf{h}$ is the effective one-body Hamiltonian (or Fock) matrix, and $\mu$ is the chemical potential. The bold-faced fonts are used to denote matrix representation. Including $\mu$ ensures the correct thermal average of the electron number. 
Detailed derivations are presented in Appendix~\ref{sec:apdx_thermo}.

\begin{figure}[t!]
    \centering
    \includegraphics[width=\linewidth]{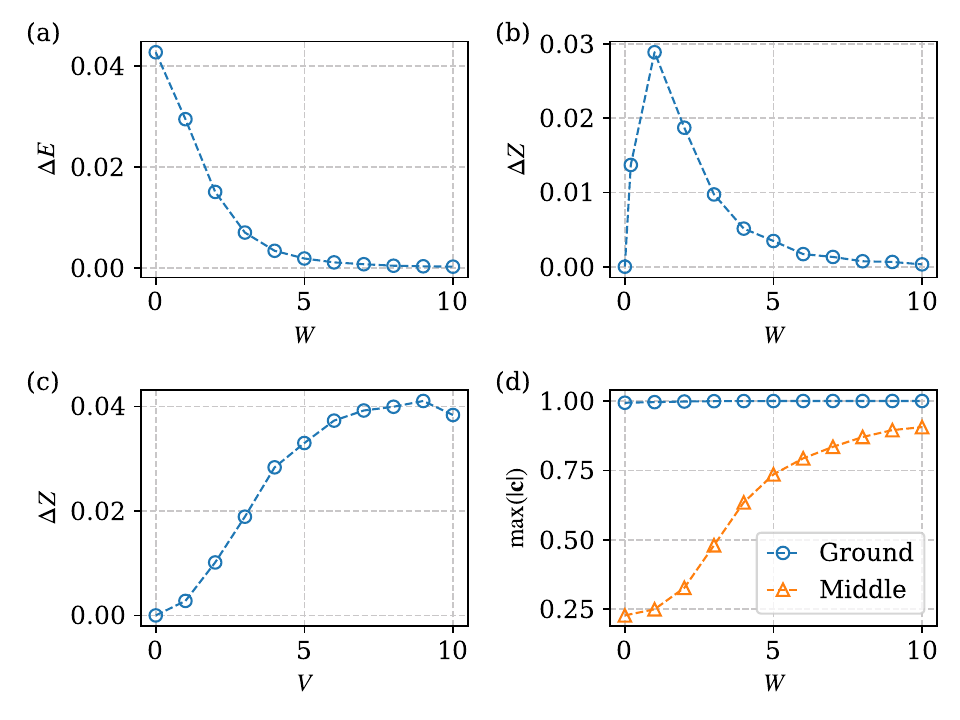}
    \caption{
    Comparison between Hartree-Fock (HF) and full configuration interaction (FCI) solutions for the disordered system at various disorder strengths $W$. (a) Ground-state energy discrepancy $\Delta E = E_{\mathrm{HF}} - E_{\mathrm{FCI}}$ as a function of $W$. 
    (b) The complex polarization magnitude discrepancy $\Delta Z = Z_{\rm HF} - Z_{\rm FCI}$ as a function of $W$. 
    (c) $\Delta Z$ as a function of the two-body interaction strength $V$ at $W = 5$. 
    (d) The maximum overlap between a FCI eigenstate and HF eigenstates. "Ground" refers to the FCI ground state, and "Middle" indicates the eigenstate at the center of the FCI energy spectrum.
     }
    \label{fig:compare_hf_fci}
\end{figure}

\subsection{Interacting formulation\label{sec:iform}}
A correlated state $|\Psi\rangle$ can be expressed as a linear combination of Slater determinants $\{|\phi_p\rangle\}$, i.e., $|\Psi\rangle = \sum_p {c_p}|\phi_p\rangle$. Consequently, $\langle \Psi|Z|\Psi\rangle$ becomes a weighted sum of overlaps between all possible pairs of Slater determinants. 
\eqsplit{
\langle \Psi|Z|\Psi\rangle &= \sum_{pq}c^*_p c_q \langle \phi_p |Z|\phi_q\rangle \\
&= \sum_{pq}c^*_p c_q \det[\mathbf{C}_{p,\mathrm{occ}}^\dagger \mathbf{Z} \mathbf{C}_{q,\mathrm{occ}}].
}
At finite temperatures, given the eigenstates of $H$ denoted as $\{|\Psi_j\rangle\}$, the thermal average of $Z$ is evaluated as
\eqsplit{\label{eq:thermal_correlate}
\langle Z\rangle (\beta) &= \frac{\sum_j e^{-\beta \varepsilon_j}\langle \Psi_j|  Z|\Psi_j\rangle}{\sum_j e^{-\beta \varepsilon_j}\langle \Psi_j|\Psi_j\rangle} \\
&= \frac{\sum_j e^{-\beta \varepsilon_j}\sum_{p q}c^*_{j,p}c_{j,q} \langle Z\rangle_{pq}}{\sum_j e^{-\beta \varepsilon_j}\sum_{p}|c_{jp}|^2},
}
where $\{\varepsilon_j\}$ are energies of the eigenstates, and $\langle Z\rangle_{pq} = \det[\mathbf{C}_{p,\mathrm{occ}}^\dagger \mathbf{Z} \mathbf{C}_{q,\mathrm{occ}}]$. Eq.~\eqref{eq:thermal_correlate} accounts for the canonical ensemble. For the grand canonical ensemble, one considers all possible electron numbers and introduces a chemical potential $\mu$ to ensure the correct thermal average of the electron number.

\begin{figure}[t!]
    \includegraphics[width=\linewidth]{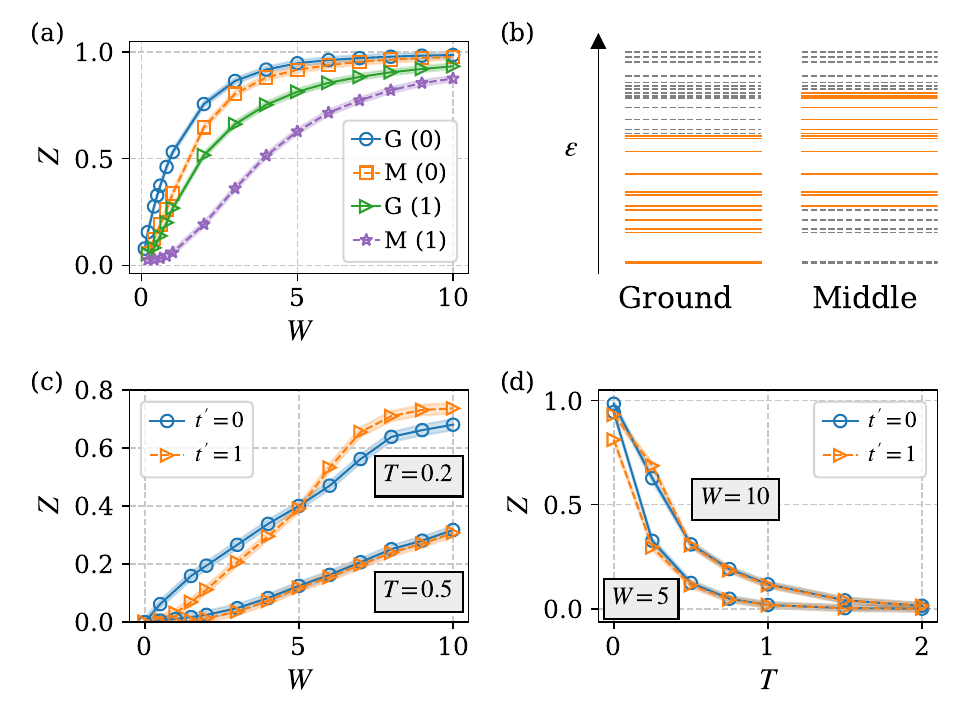}
     \caption{Complex polarization magnitude evaluated by the Hartree-Fock method. (a) Ground-state  (G) and the middle-state (M) result as a function of disorder strength $W$. Numbers in parentheses indicate $t'$ values. (b) Visualization of occupied orbitals for the ground and the middle states, respectively. (c) Finite-temperature results of $Z$ values as a function of disorder strength $W$. (d) $Z$ values as a function of temperature $T$.
      }
    \label{fig:zval_hf}
\end{figure}

\section{Results and Discussion\label{sec:results}}
We investigate a spinless fermionic chain characterized by disordered on-site potential, a setup previously explored in the context of many-body localization (MBL).  We employ periodic boundary conditions (PBC) throughout the analysis.
The Hamiltonian is given by
\eqsplit{\label{eq:ham_disorder}
H &= \sum_{\langle i, j\rangle} \left[-t\left(a^\dagger_i a_j + \text{h.c.}\right) + V  n_i n_j\right] \\
& - t'\sum_{\llangle i, j\rrangle} \left(a^\dagger_i a_j + \text{h.c.}\right) 
+ \sum_i w_i n_i,
}
where $\langle i, j\rangle$ and $\llangle i, j\rrangle$ denote nearest and next-nearest neighbor sites, respectively, and $n_i = a^\dagger_i a_i$ is the number operator. 
The hopping amplitudes $t$ and $t'$ promote electron hopping among nearest and next-nearest neighbors, respectively, while the two-body interaction term with strength $V > 0$ discourages adjacent occupations.
Random on-site potentials $\{w_i\}$ obey uniform distribution between $-W$ and $W$, introducing a disordered chemical potential to the lattice, with $W$ signifying the disorder strength. 
This disorder inhibits electron mobility and induces MBL. By convention, we set $t = 1$. We confine our studies to the half-filled regime, where the number of electrons equals half the number of sites.

In this study, we focus on the magnitude of $Z$, i.e., $|Z|$. For simplicity, we use $Z$ to represent $|Z|$ in the following discussions. 
We chose the origin of the position to be the first site with $x_0 = 0$ and limited the lattice to contain $4n + 2$ sites.
This choice avoids dealing with a zero band gap associated with a half-filled lattice of $4n$ sites. Additionally, in the finite-temperature simulations, we set the Boltzmann constant $k_B = 1$. The simulations are performed with a homemade Python package\cite{Sun2024Compol} based on \textsc{PySCF}\cite{Sun2018PySCF, Sun2020Recent}.

\subsection{Assessment of the mean-field approximation}
In the weak-interaction regime, the Hartree-Fock (HF) method typically yields accurate solutions. 
In the system studied in this work, the accuracy of HF solutions is bolstered in scenarios of strong disorder, where the one-body Hamiltonian becomes predominant. 
This insight draws inspiration from the work of Bera et al.~\cite{Bera2015Many}, who posited that natural occupation numbers~\cite{natural_occ_notes} could serve as indicators for MBL, manifesting a step-like pattern at high disorder strengths.
However, we suggest that the observed step-like pattern emerges primarily because, at high $W$, the one-body Hamiltonian prevails, rendering the system's eigenstates nearly identical to single Slater determinants (HF solution), whose natural occupation numbers are always either $0$ or $1$.
Therefore, the step-like distribution of natural occupation numbers is not clearly related to MBL. Furthermore, recent studies by Huang and colleagues~\cite{Huang2023Interaction} on the accuracy of a modified mean-field theory for the interacting GPD model~\cite{Ganeshan2015Nearest} lend additional support to our observations.

In Fig.~\ref{fig:compare_hf_fci}, we compare the HF and FCI solutions for a $14$-site disordered chain, fixing $t'=0$ and $V=1$ in accordance with the settings used in Ref.~\citep{Bera2015Many}. 
The displayed results are obtained from averaging over $1,000$ random samples. 
Fig.~\ref{fig:compare_hf_fci} (a) and (b) present the differences in energy ($\Delta E$) and complex polarization magnitude ($\Delta Z$), respectively, between the HF and FCI ground states.  
Both $\Delta E$ and $\Delta Z$ exhibit rapid declines as the disorder strength $W$ increases.
The small values of $\Delta Z$ at low $W$ are attributed to the near-zero values of $Z$. 
Fig.~\ref{fig:compare_hf_fci} (c) explores how the ground-state $Z$ varies with the two-body interaction strength $V$, under the conditions of $t' = 0$ and $W = 5$. 
In the domain of weak interactions, the discrepancy between the HF and FCI solutions is minimal. 
The observed reduction in $\Delta Z$ at $V = 9$ is, once again, a consequence of the small $Z$ values in this regime.

\begin{figure}[t!]
    \centering
    \includegraphics[width=\linewidth]{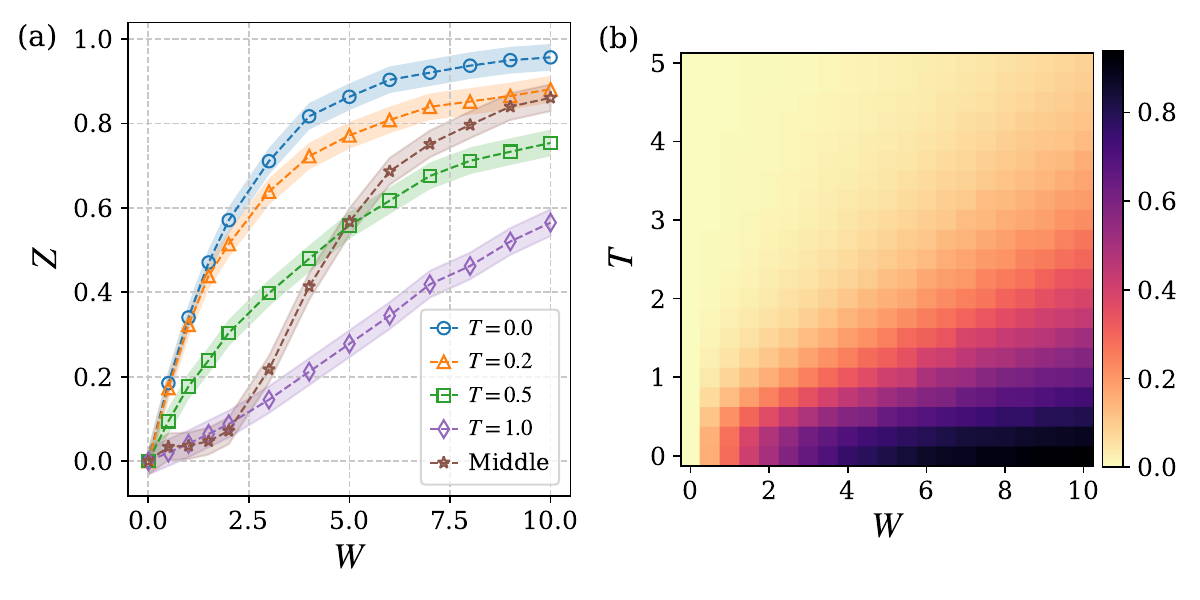}
    \caption{Complex polarization magnitude evaluated by the full configuration interaction (FCI). (a) $Z$ as a function of $W$ at various temperatures, with the shade representing the error bar due to random sampling. The $Z$ values for the middle state are also presented. (b) Heat map of $Z$ values in relation to temperature $T$ and disorder strength $W$.
    }
    \label{fig:zval_fci}
\end{figure}

To determine the accuracy of HF solutions for excited states, we analyze the normalized FCI vector coefficients $\mathbf{c}$. 
Our FCI calculations are based on the molecular orbitals (MOs) generated by HF. 
Hence, the elements of $\mathbf{c}$ represent the overlaps between all possible HF states (Slater determinants) and the FCI eigenstate, i.e., $c_i = \langle \Phi^{\mathrm{HF}}_i|\Psi^{\mathrm{FCI}}\rangle$. 
A large value of $\max(|\mathbf{c}|)$ indicates a predominant single Slater determinant within the FCI solution and, thus, high HF accuracy. 
In Fig.~\ref{fig:compare_hf_fci} (d),  we focus on the FCI ground state and an eigenstate located in the middle of the FCI energy spectrum (hereafter referred to as the "middle state") to track the behavior of $\max(|\mathbf{c}|)$ as $W$ increases. 
The two increasing curves affirm the precision of HF solutions for both ground and excited states under sufficiently high disorder strength $W$.
This result also suggests further investigation into the potential simplification of MBL to Anderson localization under strong disorder conditions, a question that extends beyond the scope of this study.

\subsection{HF and FCI simulations} 
We first analyze the electron localization of a $30$-site chain evaluated with the HF method, where the results are averaged over $N_{\mathrm{rep}} = 5,000$ random samples, shown in Fig.~\ref{fig:zval_hf}. 
The shading around the curves indicates the error bars, calculated from the worst-case sampling error $1/\sqrt{N_{\mathrm{rep}}}$, assuming the maximum single-shot error for $Z$ is $1$.
Fixing $V = 1$, we explore scenarios with ($t'=1$) and without ($t'=0$) next-nearest-neighbor hopping. 
Fig.~\ref{fig:zval_hf} (a) verifies that full localization can be attained in both ground and excited states, where we picked the middle state to showcase the excited states.
The occupied MOs for the two HF states are shown as the orange solid lines in Fig.~\ref{fig:zval_hf} (b), where the MOs are sorted according to their energy levels $\varepsilon$. The inclusion of next-nearest-neighbor hopping discourages electron localization. 
In this study, the middle states used in the HF and FCI calculations are not identical. 
The middle state for the FCI calculation is selected from the center of the eigenenergy spectrum, whereas the HF middle state is constructed by selecting the central set of MOs to form the Slater determinant.
Despite these differences in their construction, the two states are expected to be closely similar.

Proceeding to the finite-temperature results depicted in Fig.~\ref{fig:zval_hf} (c), we observe a slower growth of $Z$ with respect to $W$, with full electron localization absent even at very high $W$ values.
This indicates that thermal fluctuations diminish the impact of disorder, making the system's response to changes in $W$ less pronounced.
A notable crossover at $T=0.2$ between the cases with ($t'=1$) and without ($t'=0$) next-nearest-neighbor hopping suggests the influence of frustration.
In Fig.~\ref{fig:zval_hf} (d), $Z$ as a function of $T$ again displays a crossover around $T = 0.25$ at $W = 10$, underscoring the nuanced dynamics introduced by temperature and hopping interactions. 


\begin{figure}[t!]
    \centering
    \includegraphics[width=\linewidth]{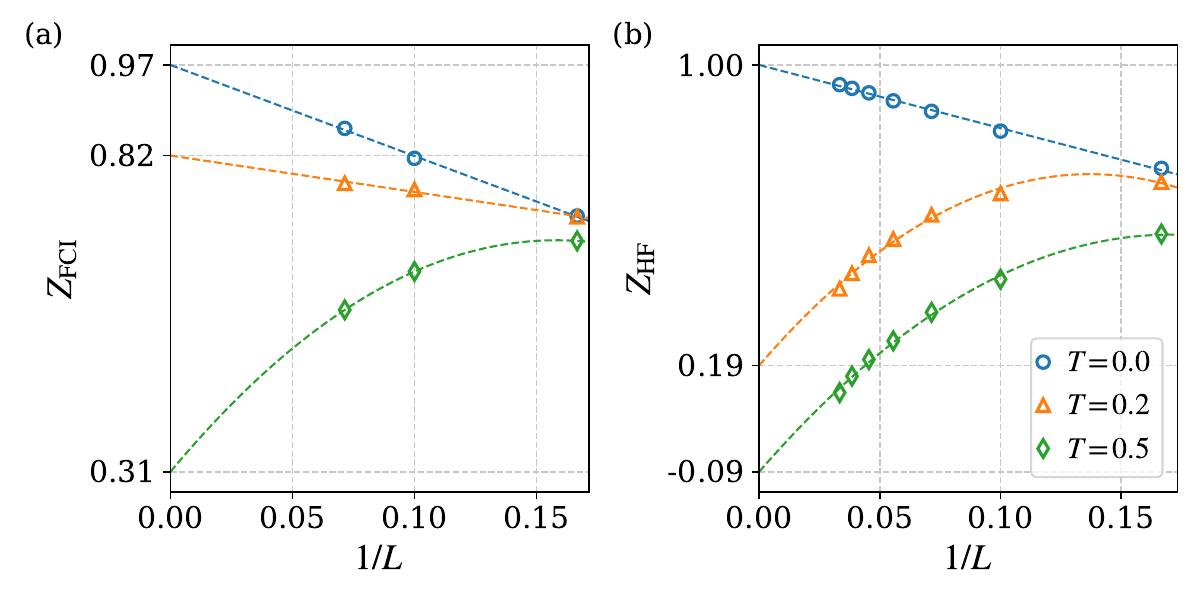}
    \caption{Extrapolation of the complex polarization magnitude $Z$ in relation to the number of sites $L$ at $W = 5$, $V = 1$ and $t'=0$. (a) FCI solutions. (b) HF solutions. Dashed lines indicate the function fitting of the discrete data.
    }
    \label{fig:extrapolation}
\end{figure}

Fig.~\ref{fig:zval_fci} displays the FCI results for the electron localization. Unlike the finite-temperature HF simulations, where the grand-canonical statistics are used, finite-temperature FCI adopted the canonical ensemble picture. 
Compared to the HF simulations, the system under the canonical statistics experiences less thermal effect. Fig.~\ref{fig:zval_fci} (a) profiles $Z$ as a function of $W$ at various temperatures and for different eigenstates (ground state and middle state). 
The simulations are performed on a $14$-site chain with $N_{\mathrm{rep}} = 1,000$ random samples, fixing $V = 1$ and $t' = 0$. At $T = 0.2$, the curve behaves like the ground state, although the maximum $Z$ value reached is slightly lower than the full localization value. As the temperature increases, the curves display linearly, similar to the HF curves. Contrary to the high-temperature curves, the $Z$ value of the middle state approaches near full localization at sufficiently high $W$ values. Fig.~\ref{fig:zval_fci} (b) presents a heatmap plot of $Z$ as a function of both disorder $W$ and temperature $T$. The simulations are performed on a small $10$-site chain with $V = 1$ and $t' = 0$. The clear boundary in the heat map further confirmed the absence of full electron localization at high temperatures. 


\subsection{Thermodynamic limit} 
To mitigate finite-size effects, we extrapolate the above simulations to the thermodynamic limit (TDL), setting $t' = 0$, $V = 1$, $W = 5$ and $N_{\mathrm{rep}} = 1,000$ across all simulations. 
We plot $Z$ against $1/L$ in Fig.~\ref{fig:extrapolation}. The FCI simulations follow canonical statistics, whereas HF simulations follow grand canonical statistics. 
The extrapolation reveals distinct behaviors for ground-state and finite-temperature conditions. At the ground state or very low temperatures, $Z$ linearly increases towards a higher value as $L$ approaches infinity, fitting the data with $Z(L) = a + b/L$. Conversely, at higher temperatures, $Z$ trends towards a lower value in a polynomial manner, fitting the data with $Z(L) = a + b/L + c/L^2$. 
Despite expectations of canonical and grand canonical solutions becoming equivalent at TDL, slight differences are observed, especially at lower temperatures where the system exhibits stronger correlations and HF simulations are less accurate. 
Additionally, the limitation to only three FCI data points may impact the precision of extrapolation. 
Integrating the heatmap from Fig.~\ref{fig:zval_fci} (b) with the extrapolation results, we anticipate a more defined boundary between localized and delocalized states in the heatmap at TDL.

\section{Conclusions\label{sec:conclusion}}
This study provided a comprehensive analysis of electron localization within disordered quantum systems, exploring both individual eigenstates and thermal states, utilizing complex polarization as the theoretical indicator.
We found that full electron localization can be achieved in individual eigenstates, while only partial localization is possible at high temperatures, despite strong disorder.
Our assessment of the Hartree-Fock method confirmed its accuracy for the system under study in conditions of high disorder.
However, for low-temperature simulations or high excited states within the weak to moderate disorder regime, more accurate methods or mean-field methods targeting the excited states, e.g., $\Delta$-SCF\cite{Jones1989Density, Gavnholt2008Delta}, are necessary. 
The methodologies and insights gained from this work offer valuable perspectives for characterizing other disordered quantum systems and could inform the discovery of disordered materials for applications like robust quantum memory.

\begin{acknowledgements}
We thank Garnet Chan, Sandro Sorella, Raffaele Resta, Huanchen Zhai, Xing Zhang, and Zijian Zhang for their insightful discussions. 
\end{acknowledgements}

\appendix
\section{Evaluating $\langle Z\rangle$ with a single Slater determinant\label{sec:apdx_slater_thouless}}

A Slater determinant $|\Phi\rangle$ can be written as the following second-quantized form
\eqsplit{
|\Phi\rangle = \prod_{p=1}^{N} d_p^\dagger |-\rangle,
}
where $N$ is the number of electrons, $d^\dagger_p$ is the creation operator on the $p$th molecular orbital, and $|-\rangle$ is the vacuum state. $d^\dagger_p$ is expressed as a linear combination of the creation operators on the site basis $\{a^\dagger_\mu\}$, 
\eqsplit{
d^\dagger_p = \sum_{\mu=1}^L C_{\mu p} a^\dagger_\mu , 
}
where $L$ is the total number of sites, and columns of the unitary matrix $\mathbf{C}$ are the molecular orbital (MO) coefficients. The first $N$ columns of $\mathbf{C}$ correspond to the occupied orbitals. Thus, the Slater determinant can be rewritten as 
\eqsplit{
|\Phi\rangle &= \prod_{p=1}^{N} \left(\sum_{\mu=1}^L C_{\mu p} {a}^\dagger_\mu\right)|-\rangle, \\
&= \sum_{1\leq \mu_1 < \cdots < \mu_N \leq L} \det[D_{\mu_1,  \cdots, \mu_N}] {a}^\dagger_{\mu_1}\cdots {a}^\dagger_{\mu_N}|-\rangle,
}
where $\{1\leq \mu_1 < \cdots < \mu_N \leq L\}$ represent all possible ${L \choose N}$ combinations of site indices, and the $N\times N$ matrix $D_{\mu_1,  \cdots, \mu_N}$ has the form
\eqsplit{{D}_{\mu_1, \mu_2, \cdots, \mu_N} = 
\begin{pmatrix}
    C_{\mu_1 1} & C_{\mu_1 2} &\cdots& C_{\mu_1 N} \\
    C_{\mu_2 1} & C_{\mu_2 2} &\cdots& C_{\mu_2 N} \\
                & & \ddots  &  \\
    C_{\mu_N 1} & C_{\mu_N 2} &\cdots& C_{\mu_N N}            
\end{pmatrix}.
}

The complex polarization is written as
\eqsplit{
{Z} = \exp\left(\sum_{\mu=1}^L {i\frac{2\pi }{L} x_\mu {n}_\mu} \right) = \prod_{\mu=1}^L \exp\left({i\frac{2\pi }{L} x_\mu {n}_\mu}\right),
}
where $n_\mu = a^\dagger_\mu a_\mu$ is the number operator on site-$\mu$. Therefore, 
\eqsplit{
{Z} \prod_{j=1}^N{a}^\dagger_{\mu_j}|-\rangle 
= \prod_{j=1}^N 
\exp\left(i\frac{2\pi}{L} x_{\mu_j}\right) 
 {a}^\dagger_{\mu_j}|-\rangle.
}

Let $\tilde{a}_\mu = \exp(i\pi x_\mu/L) a_\mu$, applying ${Z}$ onto $|\Phi\rangle$, we get
\eqsplit{\label{eq:z_phi}
{Z} |\Phi\rangle = &\sum_{1\leq \mu_1 < \cdots < \mu_N \leq L} \det[{D
}_{\mu_1,  \cdots, \mu_N}]
\tilde{a}^\dagger_{\mu_1}\cdots \tilde{a}^\dagger_{\mu_N}|-\rangle\\
=& \prod_{p=1}^{M} \tilde{d}^\dag_p |-\rangle = |\tilde{\Phi}\rangle,
}
where $\{\tilde{d}\}$ is another set of MOs,
\eqsplit{
\tilde{d}^\dag_p =  \sum_{\mu=1}^L {C}_{\mu p} \tilde{a}^\dagger_\mu =  \sum_{\mu=1}^L  e^{i\frac{2\pi }{L} x_\mu }C_{\mu p}{a}^\dagger_\mu = \sum_{\mu=1}^L \tilde{C}_{\mu p} {a}^\dagger_\mu .
}
Therefore, applying $Z$ onto $|\Phi\rangle$ results in another Slater determinant $|\tilde{\Phi}\rangle$ with MO coefficient matrix $\tilde{\mathbf{C}} = \mathbf{ZC}$, where 
\eqsplit{
\mathbf{Z} = \text{diag}\left[e^{i\frac{2\pi }{L} x_1}, e^{i\frac{2\pi }{L} x_2}, \cdots, e^{i\frac{2\pi }{L} x_L}\right].
}
Therefore,
\eqsplit{
\frac{\langle \Phi |{Z}|\Phi\rangle}{\langle \Phi |\Phi\rangle} = \frac{\langle \Phi |\tilde{\Phi}\rangle}{\langle \Phi |\Phi\rangle} = \frac{\det\left(\mathbf{C}^\dag_{\rm occ}\mathbf{Z}\mathbf{C}_{\rm occ}\right)}{\det\left(\mathbf{C}^\dag_{\rm occ}\mathbf{C}_{\rm occ}\right)},
}
where columns of $\mathbf{C}_{\rm occ}$ are the MO coefficients of the occupied orbitals.

\section{Evaluating $\langle Z(\beta) \rangle$ with thermofield theory\label{sec:apdx_thermo}}
A thermal state (or mixed state) can be derived by the partial trace of a pure state, called state purification. A textbook approach to state purification is to couple the original system $\mathcal{A}$ with an ancillary system ${\mathcal{B}}$. Given an orthonormal basis $\{|j_A\rangle\}$ in $\mathcal{A}$, we introduce an identical copy $\{|j_B\rangle\}$ in ${\mathcal{B}}$, and the purified state at $\beta = 1/k_BT$ is
\eqsplit{\label{eq:apdx_purify_state}
|\Psi(\beta)\rangle = e^{-\beta H/2}|\Psi(0)\rangle = e^{-\beta H/2} \sum_j |j_A\rangle |j_B\rangle,
}
where $H$ is the original Hamiltonian and only operates on $\mathcal{A}$. Hence, a more rigorous expression of $e^{-\beta H/2}$  is $e^{-\beta (H_A\otimes I_B)/2}$. For simplicity, we remember that physical operators only apply to states in the original system $\mathcal{A}$ and drop $I_B$ in the following. The simple summation form of $|\Psi(0)\rangle = \sum_j |j_A\rangle |j_B\rangle$ is because, at infinite temperature, all eigenstates have equal weights. Note that we do not require $|\Psi(0)\rangle$ to be normalized. 

The thermal average of $Z$ can be written as the expectation value with $|\Psi(\beta)\rangle$ 
\eqsplit{\label{eq:apdx_Z_purify_state}
\langle Z\rangle(\beta) = \frac{\langle \Psi(\beta)|Z|\Psi(\beta)\rangle}{\langle \Psi(\beta)| \Psi(\beta)\rangle}.
}
With mean-field approximation, Eq.~\eqref{eq:apdx_purify_state} and Eq.~\eqref{eq:apdx_Z_purify_state} have much simpler forms, where the MO coefficient of the thermofield state at $\beta = 0$ is a $2L\times L$ matrix
\begin{align}
    \bar{\mathbf{C}}(0) = \begin{bmatrix} \mathbf{I} \\ \mathbf{I} \end{bmatrix},
\end{align}
where the first $L$ rows correspond to the physical system $\mathcal{A}$ and the last $L$ rows correspond to the ancillary system $\mathcal{B}$.

The corresponding matrix representations of the one-body Hamiltonian $\bar{h}$ and the position operator $\bar{x}$ are
\eqsplit{
\bar{\mathbf{h}} = \begin{bmatrix} \mathbf{h} & \mathbf{0} \\ \mathbf{0} & \mathbf{0} \end{bmatrix}, \quad 
\bar{\mathbf{x}} = \begin{bmatrix} \mathbf{x} & \mathbf{0} \\ \mathbf{0} & \mathbf{0} \end{bmatrix}.
}

One can evaluate the matrix forms of thermofield complex polarization operator $\bar{Z} = \exp\left(i\frac{2\pi}{L}\bar{x}\right)$ and the density matrix $\bar{\rho} = \exp\left[-\beta(\bar{h} - \mu)\right]$ as

\eqsplit{
\bar{\mathbf{Z}} = \begin{bmatrix} \mathbf{Z} & \mathbf{0} \\ \mathbf{0} & \mathbf{I} \end{bmatrix}, 
\quad 
\bar{\bm{\rho}} = \begin{bmatrix} \bm{\rho} & \mathbf{0} \\ \mathbf{0} & \mathbf{I} \end{bmatrix}.
}
Therefore, with $\Phi(\beta)$ being the uncorrelated thermofield state,
\eqsplit{
\langle Z\rangle(\beta) &= \frac{\langle \Phi(\beta)|Z|\Phi(\beta)\rangle}{\langle \Phi(\beta)| \Phi(\beta)\rangle}\\
&=\frac{\det[\bar{\mathbf{C}}(0)^\dagger \bar{\mathbf{Z}} \bar{\bm{\rho}}(\beta) \bar{\mathbf{C}}(0)]}{\det[\bar{\mathbf{C}}(0)^\dagger \bar{\bm{\rho}}(\beta) \bar{\mathbf{C}}(0)]} \\
&= \frac{\det[{\mathbf{Z}} \bm{\rho}(\beta) +\mathbf{I}]}{\det[\bm{\rho}(\beta) +\mathbf{I}]}.
}
Thus we derived the non-interacting formulations provided in Section~\ref{sec:niform}.

\newpage
\bibliography{references}

\begin{thebibliography}{41}
\bibitem [{nat()}]{natural_occ_notes}%
  \BibitemOpen
  \href@noop {} {\bibinfo {title} {{Natural occupation numbers are the eigenvalues of the one-particle reduced density matrix (1RDM). The corresponding eigenvectors are called natural orbitals }}}\BibitemShut {NoStop}%
\end{thebibliography}

\end{document}